\documentclass[twocolumn,trackchanges]{aastex701}

\usepackage{color}
\usepackage[normalem]{ulem}
\definecolor{darkolivegreen}{rgb}{0.33, 0.42, 0.18}
\definecolor{darkpastelgreen}{rgb}{0.01, 0.75, 0.24}

\newcommand{\vunit}{\mbox{~km~s$^{-1}$}}
\newcommand{\Msun}{\mbox{\,$M_\odot$}}
\newcommand{\Rsun}{\mbox{\,$R_\odot$}}

\begin{document}

\title{Evidence for the merger hypothesis in V4332 Sgr: a low 
$^{12}$C/$^{13}$C ratio and multiple outbursts}

\author[sname='Banerjee']{D. P. K. Banerjee}
\affiliation{Physical Research Laboratory, Navrangpura,  Ahmedabad, Gujarat 380009, India}
\email[show]{dpkb12345@gmail.com}  

\author[sname='Evans']{A. Evans} 
\affiliation{Astrophysics Research Centre, Lennard Jones
Laboratory, Keele University, Keele, Staffordshire,  ST5 5BG, UK}
\email{a.evans@keele.ac.uk}

\author[sname='Varricatt']{Watson P. 
Varricatt} 
\affiliation{UKIRT Observatory, Institute for Astronomy, 640 N. Aohoku 
Place, Hilo, HI 96720, USA}
\email{w.varricatt@ukirt.hawaii.edu}

\author[sname='Ashock']{N. M. Ashok}
\affiliation{Physical Research Laboratory, Navrangpura,  Ahmedabad, Gujarat 380009, India}
\email{ashoknagarhalli@gmail.com}

\correspondingauthor{D. P. K. Banerjee}

\begin{abstract}
Following the detections of the first extragalactic  ``Luminous Red Nova'' (LRN) M31 RV in 1989, and its
first Galactic counterpart
V4332~Sgr in 1994, there have been many discoveries of similar,
or closely related, objects. They are important because they
bridge the luminosity gap between the brightest novae and
supernovae, a largely unexplored parameter space. The cause of 
eruptions in LRNe is still unclear, a stellar merger being
the most favored mechanism. However, barring V1309~Sco, there has
been no direct evidence for a merger in the other objects. Here we present strong evidence that V4332~Sgr was a merger
event. High resolution infrared observations of the CO fundamental band show an unusually small $^{12}$C/$^{13}$C
ratio of $3.5\pm1$. This indicates that a violent event had occurred, whose effects penetrated deep enough to allow CNO
cycle processed $^{13}$C in the inner H burning shell to be
brought to the surface. We rule out planetary ingestion, and
propose that the eruption was due to a merger between
V4332~Sgr and a companion star. 
It is shown that V4332~Sgr was likely surrounded by
an edge-on disk before its eruption.  If this disk was a
flattened common envelope containing V4332~Sgr
and a companion star, then a merger scenario would
not be inconsistent.
Furthermore, V4332~Sgr had multiple outbursts, previously
unreported but an important piece of information, since
multiple outbursts are a trait shared by many LRNe.
\end{abstract}

\keywords{Stellar mergers (2157) ---
Stellar astronomy (1583) 
}

\section{Introduction} 

V4332~Sgr erupted in a nova-like explosion on 1994 February 24
\citep[][]{hayashi94} and reached $\sim8$ apparent magnitude at its peak
in the $V$ band \citep[][]{martini99}. However, soon after the
explosion, its spectrum evolved to a cool spectral type, 
in contrast to novae which become hotter. In a few weeks/months
it resembled a M type star, hence the name of 
``luminous red nova'' (LRN; we use the terms Red Novae and Luminous Red Novae interchangeably).
At much later stages, several molecular species were observed in
its spectrum. In particular, water ice and CO emission in the
fundamental band was seen, along with copious amounts of dust.
The object then became shrouded in a cold dusty environment
\citep{banerjee03,banerjee04a,banerjee04b,banerjee07},
which has persisted even as
recently as 2018 \citep{kaminski18}.

V4332~Sgr was the first Galactic object observed to show these
characteristics; though previously an extragalactic LRN
(M31RV in M31) had also shown a similar nova-like eruption that
progressed towards cool temperatures \citep{mould90,rich89}. 
With the subsequent outburst of another LRN V838~Mon in early
2002, it was proposed that the eruptions of these objects were
caused by stellar mergers \citep{soker03}.
However, unambiguous proof for a stellar merger came only in the
case of V1309~Sco, which erupted in 2008 September
\citep[][]{nakano08,tylenda11}. OGLE 
\citep[][]{udalski97,szymanski05}, which had monitored
the object even before the eruption, obtained a light curve
which showed a steady brightening of the system spread over
$\sim200$~days, culminating in a peak with $I = 6.8$. 
During the steady brightening, analysis of the OGLE data showed
a periodicity of two stars spiraling into each other 
\citep[][]{tylenda11}. They found that the V1309~Sco progenitor
was a contact binary with an orbital period of $\sim1.4$~day,
its period decreasing with time. The light curve of the binary
was also seen to be evolving, indicating that the system 
components were spiralling in towards a merger.

Several other LRNe have since been discovered, most of them
extragalactic, for example, NGC4490-2011OT1, M101-2015OT1, SNhunt248
\citep[e.g.,][and references therein]{pastorello19,kasliwal12}.
These are mostly more luminous ($M_V < -13$) than their Galactic
counterparts, but have many common properties. However, with the 
exception of V1309~Sco, the hypothesis that LRNe in general
are stellar mergers (or merge-burst objects) lacks direct
observational proof. These objects also have a large
spread in their outburst luminosities, making it difficult to
perceive that their outbursts have a common mechanism.
The absolute magnitudes at outburst of V3442~Sgr and
V1309~Sco were between $-6$ to $-7$ (subject to their uncertain
distances); both V838~Mon and M31RV had $M_V\simeq-10$
\citep[][]{bond03,rich89}, CK~Vul has $M_V = -12.4$
\citep{banerjee20}, while some of the extragalactic 
LRNe have even brighter $M_V$ values 
\citep[e.g.,][]{pastorello19,kasliwal12}. In terms of kinematics,
their line profiles typically show outflow velocities of a
few hundreds of \vunit, much lower than novae 
\citep[e.g.,][]{kaminski18}. In the case of V4332~Sgr 
specifically, definite proof that matter was ejected
during its eruption is seen from HST imaging 
\citep{bond18}, and from imaging in the millimeter by
\cite{kaminski18}, both of which resolve the object spatially.

In the present study, our main objective is to present fresh
evidence which strengthens the case that V4332~Sgr is indeed
a merger object. Two scenarios are examined, namely whether the
merger was between two stars, or alternatively whether a star
ingested a planet.

\section{Observations}
V4332~Sgr was observed using the 3.8-m UKIRT telescope
(project ID U/05A/12) and the Cooled Grating Spectrograph~4 
\citep[CGS4;][]{mountain90}.
CGS4 is a 1--5\micron\ grating 
spectrometer with a $256\times256$ InSb array. We used 
the 2-pixel-wide slit and CGS4's echelle grating to get
a spectral resolution $R = 18500$. The observations were
performed by offsetting the telescope to two positions
separated by $9''$ in an ABBA pattern along the slit oriented
North-South (slit angle $0^\circ$). This pattern was repeated
to increase the depth as required.

Observations were obtained at three different grating
settings; together, these cover the CO fundamental
band heads in the wavelength range 4.65--4.737\micron. Each
target observation was preceded by the observations of a 
telluric ratioing star, black body flat and an Argon arc
spectrum at the same instrument setting as for the target.
The G6V star BS 6998 was used as the ratioing star.
Table~\ref{log} gives the details of the observations.

The preliminary reduction was performed using the
facility pipeline ORACDR. The object and telluric standard
spectra were flat fielded and the dithered frames at two
positions in the ABBA set were mutually subtracted to remove
the sky and the instrumental background. The subtracted pairs
were coadded to create the reduced spectral images, which have
the spectra in the +ve and --ve beams from mutually subtracted
dithered observations. Further reduction involving optimal
extraction and averaging the spectra from the two beams
and combining them, division by the telluric standard and 
the wavelength calibration were performed using the tasks 
in the FIGARO package 
\citep{shortridge12}
of the STARLINK data reduction package \citep{giaretta04}.
As the arc spectra did not have any lines in the wavelength
ranges of our observations, airglow lines in the raw spectral 
images were used for wavelengh calibration. Fig. 2 shows 
the combined spectra from the three sets of observations.

\begin{deluxetable*}{cccccc}
\digitalasset
\tablewidth{0pt}
\tablecaption{Log of observations.\label{log}}
\tablehead{}
\startdata
Date & UT  &Central   &    Standard star   &        Number   & Total \\
YYYY-MM-YY & & wavelength (\micron) & Object & of coadds$^a$  & Integration (s)  \\ \hline
2005-07-12 &   10:00:00 & 4.695 &    BS6998   &  10 & 320  \\
         & 10:27:30  &      &   V4332Sgr &    15 & 1680  \\
2005-07-30 &  06:26:30 & 4.72    &   BS6998     & 20 &  640  \\
         &   07:15:00   &     &  V4332Sgr    & 20 &  1920 \\
2005-07-30 &  10:19:00 & 4.67    &  BS6998      & 20 &  640 \\
         &   10:27:30  &      &  V4332Sgr    & 20 & 1600  \\
\hline
\enddata
\tablecomments{$^a$1 second per frame.}
\end{deluxetable*}

\section{Results}
\subsection{An updated Light Curve for V4332 Sgr}
In a meticulous calibration of old archival plates,
\cite{kimeswenger06} showed that V4332~Sgr brightened by
over a magnitude in the blue between 1950 and 1976, 
and by a magnitude in the red from between 1950 and 1985
(although there were no intervening data in either color).
The brightening subsequent to $\sim1990$ was very rapid.
Of course, V1309~Sco brightened relatively quickly,
from $m_I = 17$ to its peak of $I = 6.8$ in about 210~days.

\begin{figure}
\centering
 \includegraphics[width=0.31\textwidth,angle=-90]{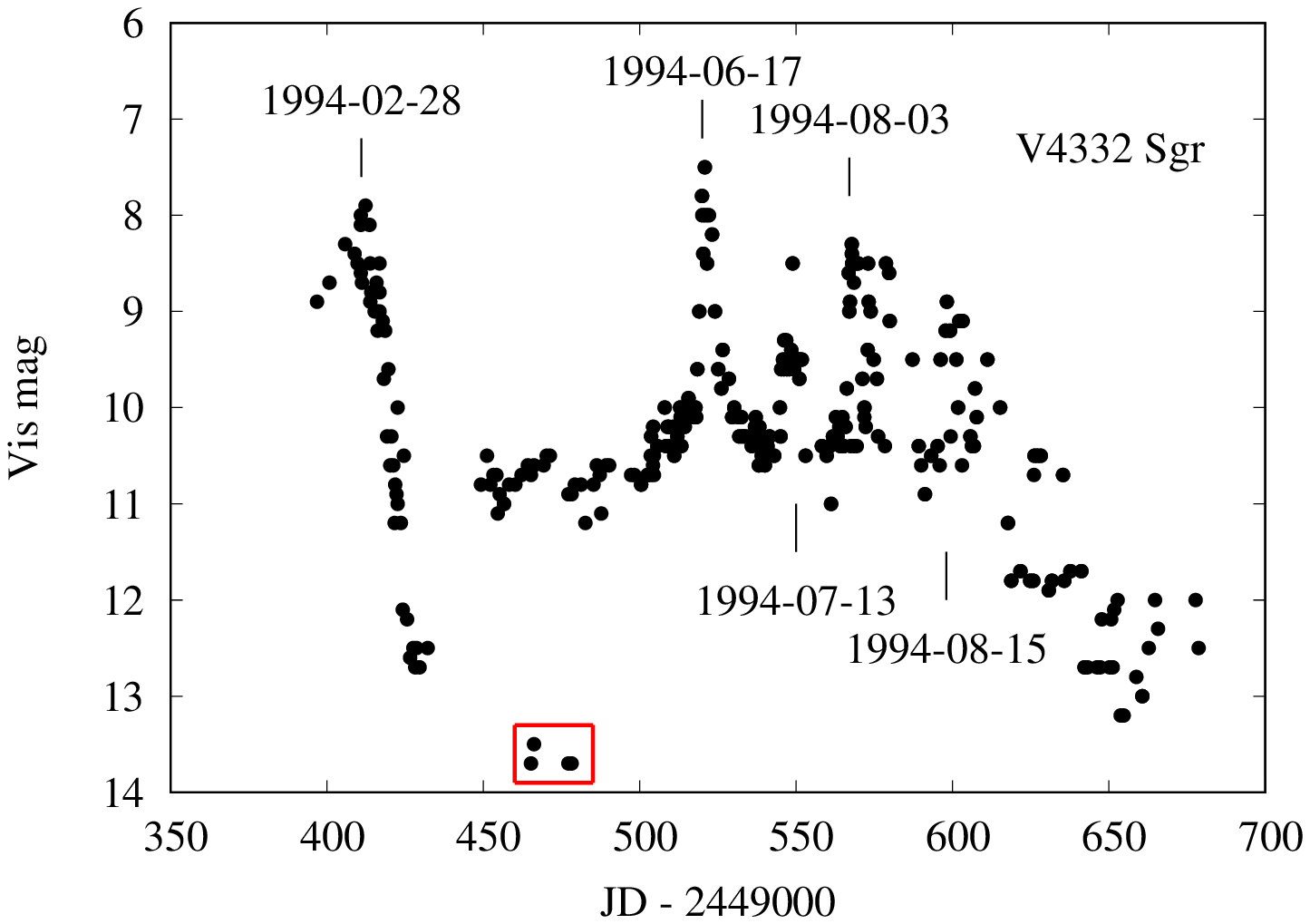}
 \includegraphics[width=0.45\textwidth]{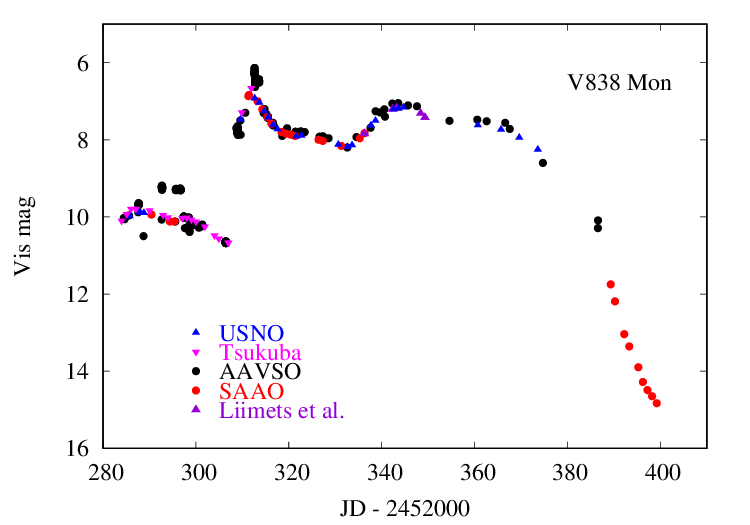}
 \includegraphics[width=0.44\textwidth]{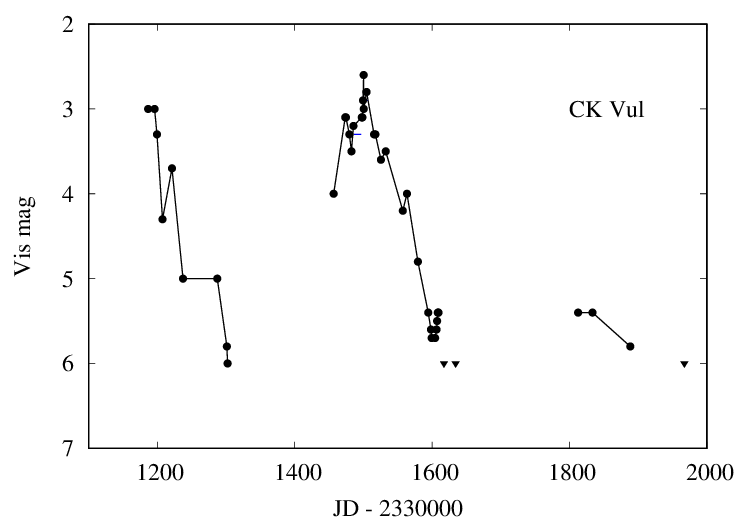}
 \caption{Top: light curve of V4332 Sgr, showing three 
 clear and two possible peaks. Data from AAVSO and AFOEV.
 See text for discussion of the three points enclosed in the red square.
Middle and bottom: V838~Mon and CK~Vul, similar objects
displaying multiple outbursts;
the data for CK~Vul are from Table~1 of \cite{shara85}
\label{LC}}
\end{figure}

The light curve immediately following the outburst is shown in 
\cite{martini99}, 
who discuss the quality of the early data (some of which
were obtained under marginal conditions). The light curve
shows a decline by two magnitudes in
$8\pm1$~days. However, 
coverage stops about 25 days after the peak, and little
attention has been paid to the evolution of the light curve
of V4332~Sgr thereafter. 
A composite light curve consisting of 
AFOEV\footnote{https://cdsarc.cds.unistra.fr/ftp/afoev/} and 
AAVSO\footnote{https://www.aavso.org/}
data, shown in Fig.~\ref{LC} 
\citep[in which the data as far as
day 440 are identical with those of][]{martini99},
demonstrates that V4332~Sgr had multiple episodes of 
re-brightening, with distinct peaks on 1994 June 17,
July 13, August 3, and possibly August 15 (where the
light curve is quite erratic). Note, however, the
discrepancy between the three points around day 460
($\sim13.6$ magnitude; contained in red square), 
and other data around this
time ($\sim10.8$ magnitude). Nonetheless,
it is very clear that V4332~Sgr underwent multiple
secondary outbursts. The peak of the second outburst, although
based on a visual estimate, was at $V\sim7.5$ and even brighter
than the primary peak at $\sim8$~mag. The multiple outbursts
establish that V4332~Sgr shared a common property with some
other known or suspected Galactic merge-burst objects like
V838~Mon or CK~Vul. Fig.~\ref{LC} illustrates this well.

\subsection{The unusually low $^{12}$C/$^{13}$C ratio in V4332~Sgr}

\begin{figure}
 \centering
 \includegraphics[width=0.5\textwidth,keepaspectratio]{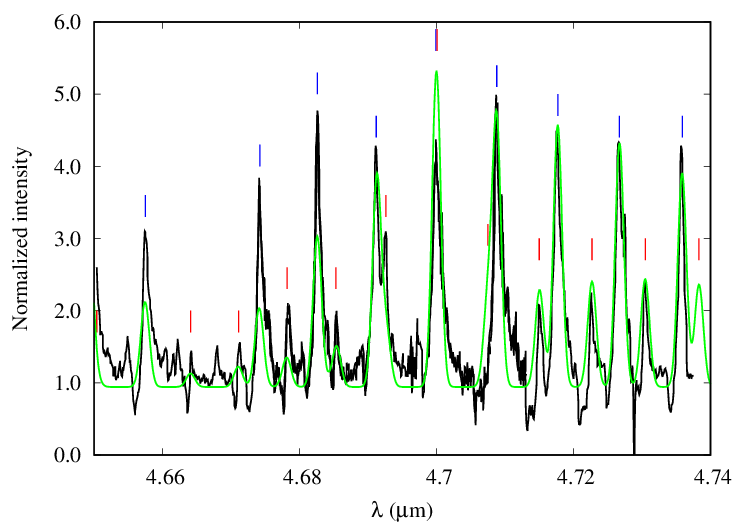}
 \caption{Observed echelle spectra of V4332~Sgr (black) with
 the model fit (green). The $^{12}$CO and $^{13}$CO rotational
 lines are marked with blue and red ticks respectively. 
 Details are given in Table~\ref{log} and in the text.
 \label{CO}}
\end{figure}

The echelle spectra of the CO fundamental are shown in 
Fig.~\ref{CO}, along with a model fit.
Details of the model, which assumes the presence of
$^{12}$C$^{16}$O and $^{13}$C$^{16}$O and optically thin, LTE emission, may be found in \cite{das09}. 
Isotopologues of CO with $^{17}$O or $^{18}$O were not
considered as there is no evidence of emission 
at the wavelengths of their rotational lines.
Reasonably good fits are found for
gas temperatures in the range 150--300~K 
\citep[as we had found by modeling the low-resolution spectra in][]{banerjee04b},
and for a $^{12}$C/$^{13}$C value of $3.5\pm1.0$.
There appear to be P-Cygni features to the CO lines, 
and there may be non-LTE effects, but modeling these is beyond
the scope of the paper. Essentially, the presence of substantial
$^{13}$CO can be assessed even from a visual examination of
the strengths of the R6, R5 and R4 lines at 4.7150\micron,
4.7227\micron\ and 4.7305\micron,
which are resolved and unblended with $^{12}$CO rotational lines.

The observed $^{12}$C/$^{13}$C value is strikingly small. 
The nature of the progenitor is uncertain, but if it
was a main-sequence star, as suggested by \cite{tylenda05},
based on the apparent magnitudes and colors of the
progenitor in archival plates, the ratio is expected to be
similar to the solar value of 89 \citep{asplund21}.
\citeauthor{tylenda05} however also discuss the possibility of
it being a giant, for which a larger distance to the
object would be needed (as we assume in Section~\ref{dc}). 
If the progenitor was a giant that had begun its ascent up
the giant branch, and had undergone a first dredge up, then the
theoretical $^{12}$C/$^{13}$C ratio should lie between 18 and 26
for low and intermediate mass
stars \citep[see Figure 6 or Table 1 of][]{karakas14}. However,
there are deviations from this range of the 
$^{12}$C/$^{13}$C ratio in open metal-rich clusters which show
values of 20, sometimes 10, and even more striking deviations are
seen in metal-poor field stars and in giants in globular clusters
\citep[][and references therein]{karakas14}.
Thus an additional mixing process has to be invoked that can connect the
hot region at the top of the H-burning shell with the convective envelope
above. Only then can can products from the CNO cycle in the shell
(e.g., $^{13}$C, $^{15}$N, $^{17}$O) be brought to the surface. As an
aside, similarly in planetary nebulae, where such small or even
smaller $^{12}$C/$^{13}$C ratios have found, 
\cite{2020ApJ...900L..31Z}
have pointed out the need for some explosive event that enhances
the mixing (and thus lowers the $^{12}$C/$^{13}$C ratio). In V4332~Sgr, we
propose that the low observed $^{12}$C/$^{13}$C makes a strong
case that a violent planet ingestion or stellar merger had taken place,
in which the merging body had plunged deep into the primary star and
caused the additional mixing between layers.

\begin{figure}
 \centering
 \includegraphics[width=0.5\textwidth,keepaspectratio]{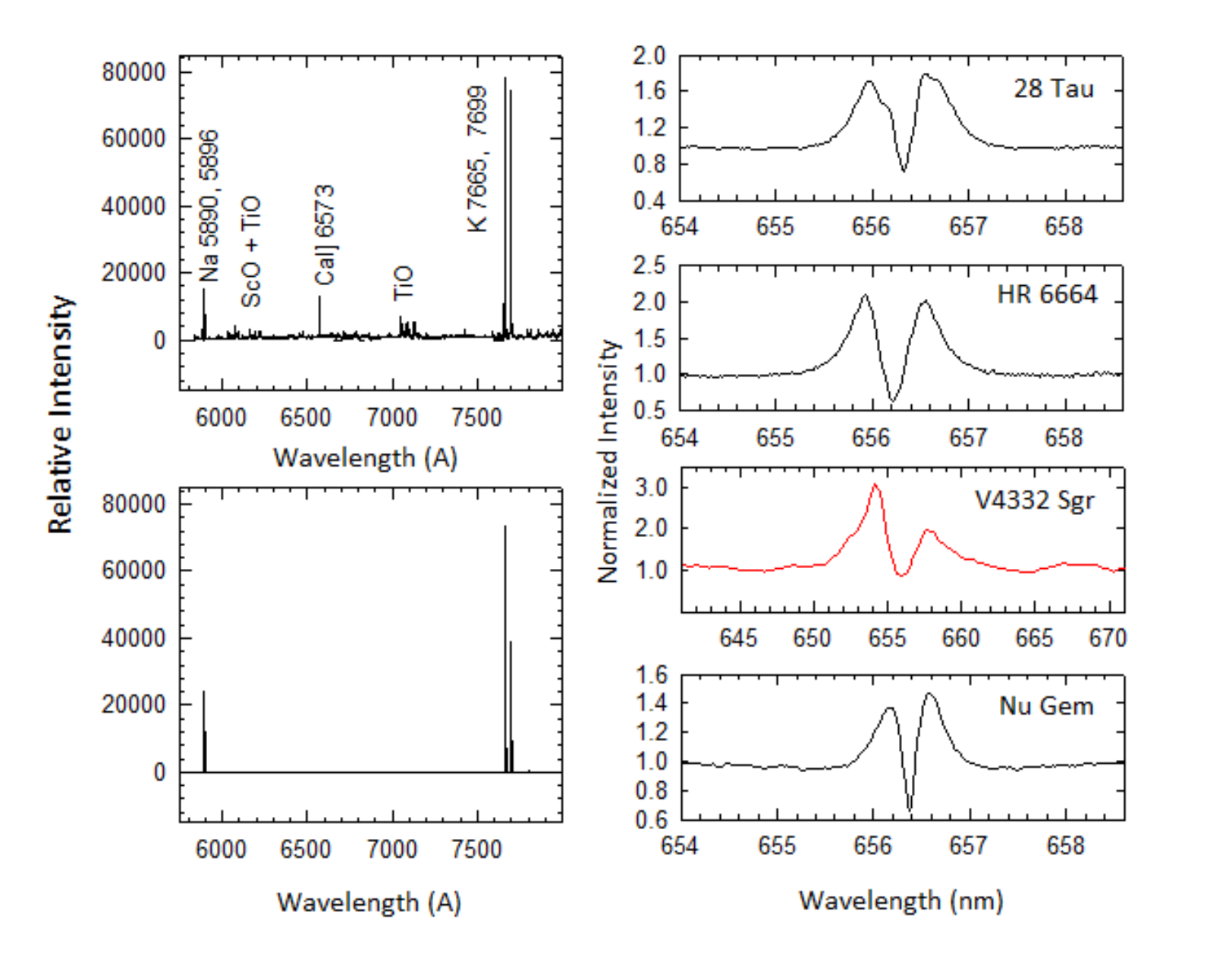}
 \caption{Top left: strong emission in the sodium and potassium
resonance lines in a small section of the V4332~Sgr spectrum taken
in 2005 from the Very Large Telescope by one of us (DPKB; PI of
program ESO-VLT 075.D-0511). These data are analyzed in 
detail by \cite{tylenda15}. The bottom panel shows a LTE
model plot to show that the unusual strength of the
K and Na lines can be explained without using enhanced abundances
(see text). Right column: H$\alpha$ profile of V4332~Sgr (red)
on 1994 March 7 comparing its deep central absorption with the shell
profiles of three Be stars \citep[from][]{silaj10} to show the
similarity in profiles and hence establish that V4332~Sgr had a disk
around it even before its 1994 eruption. See text for details.
 \label{Spectra}}
\end{figure}

Planetesimal infall into a star has been observed 
to occur \citep[][]{de23}, and we consider this possibility
in the case of V4332~Sgr for two reasons. 
First, V4332~Sgr showed unusually strong resonance
lines of K and Na in emission in a spectrum taken
10.5~years after outburst \citep{banerjee04a}; see 
Fig.~\ref{Spectra}. Na was also prominent
immediately after the 1994 outburst \citep{martini99}. 
We examined whether the strengths of the K and Na lines in 
Fig.~\ref{Spectra} were due to an
overabundance of these metals, or to physical conditions conducive to producing strong emission in these lines. 
The crust
of rocky planets have substantial K and Na which can be
released on ingestion. Na and K have also been found in the
atmospheres of exoplanets, in agreement with
theoretical predictions \citep{charbonneau02,keles19}.
Several white dwarfs have been found 
(e.g., GD 40, WD J2147-4035, WD J1922+0233), that unexpectedly
show lines of Li, Na and K, suggesting they had consumed a
rocky extrasolar object during
their past history \citep[][]{kaiser21,elms22}.

A model spectrum of V4332~Sgr was generated using the NIST 
Saha-LTE application\footnote{Available at
https://physics.nist.gov/PhysRefData/ASD/LIBS/libs-form.html} 
assuming solar abundances for the elements. We are able to
satisfactorily reproduce the observed line strengths of
Na and K, without the need of enhancing their abundances
(w.r.t. solar), by considering a low temperature gas at
$kT = 0.095$~ev ($\sim1100$~K) with an electron density in the
range $10^8$ - $10^{10}$~cm$^{-3}$. The low temperature is
consistent with the rich molecular environment
(many bands of AlO, ScO, TiO are seen). The 
electron density is also
consistent with the $n_e = 10^8 - 10^9$~cm$^{-3}$ estimated by
\cite{martini99} from \ion{O}{1} lines. With this choice of
parameters, the lines of H and He, comprising almost 99\% of
the gas, are suppressed in the model spectrum. This is consistent
with the fact that they are not seen in the observed spectrum also. Even the D1, D2 resonance lines of Rubidium at 7800\AA\
and 7948\AA\ that are observed in the spectrum, are generated
in the model in spite of the low Rb abundance assumed
($\log(N\mbox{(Rb)}) = 2.4$ on a scale with
$\log(N\mbox{(H)}) = 12$. Although the model is simple,
it shows that the
strong K and Na lines are not due to K and Na-rich material
expelled after planet ingestion. 
The above adds weight (as per Section~\ref{dc})
that the eruption was {\em not} planet ingestion.

Second, there may be circumstantial evidence
for the presence of a pre-existing disk (possibly common
envelope material) around V4332~Sgr.
Fig.~\ref{Spectra} shows the high-resolution H$\alpha$
spectrum observed by \cite{martini99} on 1994 March,
just 11 days after the outburst. It is reproduced here from the
original data archived in the ESO science data
portal\footnote{https://www.eso.org/public/science/archive/}. 
The H$\alpha$ profile shows a very deep central reversal, 
a phenomenon that is
often seen \citep[albeit not exclusively; see, e.g.,][]{lebre91} in certain Be stars known as shell stars. 
Be stars show a variety of profiles 
\citep[e.g.,][]{banerjee00,silaj10}, but the shell stars
\citep[Be stars seen with their ionized disks almost edge-on; e.g.,][]{struve31,porter03},
exhibit a deep central reversal, with the deepest part of the absorption sinking even below the continuum level.
The phenomenon involves absorption of
the photospheric flux throughout the whole disk, and hence provides information
on the radial velocity structure \citep[][]{porter03}. 
H$\alpha$ spectra of three such Be shell stars are shown in
Fig.~\ref{Spectra}, where the 
resemblance of the line
profiles between the V4332~Sgr and these Be stars is evident.
The V4332~Sgr profile can be explained if the H$\alpha$ emission from the ejecta,
created during the eruption, passes through the cold gas of a surrounding disk, which we
hypothesize to be present. This will thereby produce the central absorption in
the profile. In addition, even the continuum of the erupting star at the
H$\alpha$ wavelength can be absorbed by the disk. This will make the central
reversal fall even below the continuum level, as is observed here. 
If there is a disk, then it is always plausible to harbor a planet that can
migrate inwards and get captured by the star. On the other hand, if the disk
is the flattened remnant of the common envelope phase, it will then harbor a
companion which can spiral in due to viscous drag and end up in a merger.
We are thus proposing that the line profiles may indicate the  presence of a pre-existing disk around V4332~Sgr. This is mentioned as a possibility, because a similar deep absorption in the emission line profile could be caused by other mechanisms 
\citep[e.g., in late-type variable stars,
due to shocks in the stellar atmosphere rather than disks, as
in the case of R~Scuti;][]{lebre91}.

It should be mentioned that the presence of an edge-on disk was also suggested,
based on other lines of argument, including polarization measurements, 
by \cite{kaminski11} and \cite{kaminski13}. But while they say that the dusty
disk or disk-like envelope most probably formed during the 1994 eruption of V4332~Sgr
\citep{kaminski13}, we argue that it existed even before that.

\section{Discussion and Conclusion\label{dc}}

We estimate the minimum total energy
radiated during the outburst for the first 280 days between
days MJD 49396 to 49677 shown in Fig.~\ref{LC} to 
be {\em at least} $\sim2\times10^{46}$~erg. This was determined assuming that $ 0.35<E(B-V)<0.75$ \citep{tylenda15}, and that
the distance is at least 4~kpc 
\citep[see, e.g.,][]{kaminski18}; we further
assumed that the bolometric correction is $<0$.
We do not the Gaia estimates because Gaia EDR3 gives a negative parallax for the V4332~Sgr. In any case, the
image of the star is clearly non-stellar \citep{bond18}, 
which could well compromise the
parallax measurement.
A similar value ($10^{46}$~erg) for the total energy
radiated by M31 RV in the first 100 days was
estimated by \cite{rich89}. 

Engulfment of an Earth-like planet (mass $M_{\rm p}$) 
orbiting a star having solar mass and
radius would release an energy \citep[see, e.g.,][]{macleod18}
$\sim{G}\Msun{M}_{\rm p}/2\Rsun\sim5\times10^{42}$~ergs,
which is much smaller than the estimated total energy released.
Estimates show that this conclusion would not change even
if a Jupiter-sized planet is considered. It may be 
added that in ZTF SLRN-2020, somewhat similar to a {LRN-type}
eruption in which a planet was engulfed \citep{de23},
the total radiated energy was $\simeq6.5\times10^{41}$~erg.
This pointed to the engulfment of a planet of
fewer than roughly ten Jupiter masses by its Sun-like
host star \citep{de23}.

Hence we conclude, on the evidence of the 
$^{12}$C/$^{13}$C ratio, of the energetics of the eruption,
and of the Na/K lines,
that a stellar merger, rather than planet ingestion, is 
the more plausible explanation for the eruption of
V4332~Sgr.

\begin{acknowledgments}
We thank the referee for their constructive remarks.

UKIRT is owned by the University of Hawaii (UH) and
operated by the UH Institute for Astronomy. When the data
reported here were obtained, UKIRT was operated by the Joint
Astronomy Centre on behalf of the Science and Technology
Facilities Council of the U.K.

We acknowledge with thanks the variable star observations from the AAVSO International Database contributed by observers worldwide and used in this research. We also thank the AFOEV observers who have contributed to this
work.
\end{acknowledgments}

\facilities{UKIRT(CGS4), Gaia, ESO, AAVSO}

\software{FIGARO}

\bibliography{V4332_v13.bib}{}
 \bibliographystyle{aasjournalv7}

 \end{document}